\documentclass[conference]{IEEEtran}

\ifCLASSINFOpdf
\else
\fi

\usepackage{multirow}
\usepackage{multicol}

\usepackage{cite}
\usepackage[textsize=tiny]{todonotes}

\usepackage{listings}
\usepackage{xcolor}

\definecolor{codegreen}{HTML}{F3F5D9}
\definecolor{codegray}{HTML}{A6B0A0}
\definecolor{codegreen}{HTML}{8DA101}
\definecolor{codepurple}{HTML}{DF69BA}
\definecolor{backcolour}{HTML}{FFFBEF}
\definecolor{fgcolor}{HTML}{5C6A72}

\lstdefinestyle{mystyle}{
    backgroundcolor=\color{backcolour},
    commentstyle=\color{codegreen},
    keywordstyle=\color{codepurple},
    numberstyle=\tiny\color{codegray},
    stringstyle=\color{codegreen},
    identifierstyle=\color{fgcolor},
    basicstyle=\ttfamily\footnotesize,
    breakatwhitespace=false,
    breaklines=true,
    captionpos=b,
    keepspaces=true,
    numbers=left,
    numbersep=5pt,
    showspaces=false,
    showstringspaces=false,
    showtabs=false,
    tabsize=2
}

\begin{document}
\title{Adding MFMA Support to gem5}

\author{\IEEEauthorblockN{Marco Kurzynski}
\IEEEauthorblockA{\textit{ECE Department} \\
\textit{University of Central Florida}\\
Orlando, FL, USA \\
marco.kurzynski@ucf.edu}\\
\and
\IEEEauthorblockN{Matthew D. Sinclair}
\IEEEauthorblockA{\textit{Computer Sciences Department}\\
\textit{University of Wisconsin-Madison}\\
Madison, WI, USA \\
sinclair@cs.wisc.edu}
}

\maketitle

\IEEEpeerreviewmaketitle

\section{Introduction}
\label{sec:intro}

\IEEEPARstart{I}{n} recent years machine learning (ML) has emerged as an important application domain driving the requirements for future systems.
As DNNs get more sophisticated, their compute requirements and the datasets they are trained on continue to grow rapidly.
For example, Gholami, et al. showed that compute in the widely used Transformer networks~\cite{ShoeybiPatwary2019-megatronlm} grew 750$\times$ over 2 years~\cite{MLGrowth}, while other work projects DNN compute and memory requirements to grow by 3$\times$ per year~\cite{NaffzigerBeck2021-ryzen, JouppiKurian2023-tpuV4, JouppiYoon2021-tpuv4i}.
Given ML's growing requirements and importance in driving the requirements of system design, heterogeneous systems often add ML specific features (e.g., Matrix Core Engines or TensorCores) to improve efficiency for ML workloads.
However, given ML's voracious rate of growth and size, there is a growing challenge in performing early-system exploration to identify promising optimizations for future systems.

For example, the gem5 simulator is a popular cycle-level system simulator for studying computer hardware and systems~\cite{binkert2011gem5, LowePowerAhmad2020-gem520}.
It supports various CPUs, GPUs, and other important accelerators~\cite{GutierrezBeckmann2018-amdGem5, RoartySinclair2020-gem5GPU, RogersSlycord2020-gem5Salaam, SpencerRogers2024-gem5Salam2, YogatamaSinclair2020-multiGPU}.
Unfortunately, the gem5 simulator currently does not support modern hardware features.
For example, modern AMD and NVIDIA GPUs support specialized hardware units called Matrix Core Engines (MCEs~\cite{SchiefferDeMedeiros2024-mce}, AMD) or TensorCores~\cite{ChoquetteGiroux2018-volta, RaihanGoli2019-tc} (NVIDIA).
These specialized hardware matrix units efficiently perform various mathematical operations that are heavily utilized in state-of-the-art ML workloads.
For example, on AMD MI200~\cite{LohSchulte2023-mi250} and MI300~\cite{SmithLoh2024-mi300A} GPUs, MCEs perform Matrix Fused Multiply Add (MFMA) instructions for a variety of precisions.
However, since gem5 does not currently support MCEs, it is very difficult for users to simulate state-of-the-art workloads in gem5.
To address this shortcoming, in this work we have enhanced gem5's GPU model support to add MCEs for the MI200- and MI300-class GPUs gem5 supports.
By adding this support, our changes enable running state-of-the-art ML workloads in gem5, as well as examining how MCE optimizations impact the behavior of future systems.

\section{Background}
\label{sec:back}

The gem5 simulator is a widely used, open-source, cycle-level computer system simulator.
At its core gem5 contains an event-driven simulation engine.
On top of this simulation engine gem5 implements a large number of models for system components for CPUs (out-of-order designs, in-order designs, and others), GPUs (AMD and ARM models), accelerators~\cite{RogersSlycord2020-gem5Salaam, ShaoXi2016-gem5Aladdin}, various memories, on-chip interconnects, coherent caches, I/O devices, and many others.
Moreover, gem5 provides two modes: Syscall Emulation (SE) and Full System (FS).
SE mode simulates an application's user mode code in detail but emulates the OS instead of simulating it in detail.
Conversely, FS mode simulates both the OS and user mode code in detail, allowing users to study the interaction between the OS and architecture.

We previously enhanced and updated gem5's GPU support~\cite{GutierrezBeckmann2018-amdGem5} to add support for multi-chiplet systems~\cite{YogatamaSinclair2020-multiGPU} and ML workloads~\cite{RoartySinclair2020-gem5GPU, BruceAkram2021-gem5art} in SE mode.
As a result, ML workloads such as DNNMark~\cite{DongKaeli2017-dnnmark} and DeepBench~\cite{Narang2017-deepBench}, which call ML libraries directly (e.g., using MIOpen~\cite{KhanFultz2019-miopen} or rocBLAS~\cite{rocblas} API calls) run in gem5's SE mode.
While this represents a significant step forward, this approach is insufficient for modern high-level frameworks such as PyTorch~\cite{AnselYang2024-pytorch2, paszke2017-pytorch} and TensorFlow~\cite{tensorflow2015-whitepaper}.
Accordingly, we recently released support for running ML models in gem5's FS mode~\cite{RamadasPoremba2024-gem5GPUFS, RamadasSinclair2024-gem5GPUFS}.
These high-level frameworks utilize highly tuned libraries, which frequently utilize MFMAs, to run the high-level code at high efficiency on a given target backend (e.g., an AMD GPU).
Thus, users can now run PyTorch and TensorFlow workloads on CPU-GPU systems in gem5 using modern versions (e.g., v6) of AMD's open-source ROCm stack.
However, since ML workloads that use PyTorch and TensorFlow frequently utilize MFMA instructions, their lack of effective, validated support in gem5 hinders users' ability to run PyTorch and TensorFlow workloads in gem5.

\section{Adding MFMA instructions to gem5}
\label{sec:des}

MFMA instructions use dedicated MCEs in the SIMD units of each compute unit (CU).
In our gem5 implementation we assume 1 MCE per SIMD unit in a CU for the simulated MI200 and MI300 GPUs.
Since there are 4 SIMD units per CU by default, this corresponds to 4 MCEs per CU.
We based this design on AMD's reported MCE operations per clock~\cite{SmithLoh2024-mi300A}, as well as the number of SIMD units per CU and CUs in MI200 and MI300 GPUs.
In hardware, the MCE are separate functional units (FUs) from the existing FU for transcendentals, vector ALU operations, vector loads/stores, scalar memory operations, and other operations~\cite{amd-cdna3}.
Thus, a GPU CU can execute instructions that utilize these different FUs concurrently.

\noindent
\textbf{gem5 Code}: In gem5 our MCE/MFMA implementation is primarily located in \texttt{src/gpu-compute/compute\_unit.cc} (timing implementation) and \texttt{src/gpu-compute/scoreboard\_check\_stage.cc} (logic for issuing MFMA instructions and determining when MCEs are available).
The functional implementation of the MFMAs is located in \texttt{src/arch/amdgpu/vega/insts/instructions.hh}.

All matrix core instructions are composed of the operation $D = C + A * B$ where $D$ and $C$ are $4\times4$ matrices, and $A$ and $B$ are $4\times1$ and $1\times4$ matrices, respectively.
In AMD's Vega ISA, these instructions are of the form \texttt{V\_MFMA\_[output type]\_[$M$]X[$N$]X[$K$]\_[$B$]B\_[input type]} where $M$, $N$, and $K$ are dimensions and $B$ is the number of these matrices that are multiplied.
Different MFMA instructions implement different shapes and blocks of these operations which can be executed by a single matrix core.
The larger the block size, the smaller the dimensions of the product, and the larger the block size and dimensions, the more cycles the instruction takes to execute in the MCE (Table 27 of the MI300 ISA manual~\cite{mi300_isa}).
In gem5, we use the \texttt{NRDY\_MATRIX\_CORE} field per SIMD unit to determine when an MCE is available: the scoreboard checks that the appropriate number of cycles for the given GPU model and instruction type have passed before allowing another MFMA to execute.

While a CU is utilizing an MCE functional unit, it can perform other work within the same thread (or in other threads from different wavefronts).
For work within a thread, the work must not have any dependencies on the data produced by the MCE.
Within a thread, modern AMD GPUs implement this by requiring that either the programmer (if using inlined assembly) or the compiler (all other scenarios) identify independent work to perform.
Frequently, this takes the form of software pipelining across multiple loop iterations that utilize MFMA instructions.
However, if no independent work is available, then the compiler (or programmer) must insert NOPs (\texttt{s\_nop}) to ensure no dependent work is performed until the MFMA instruction has completed (e.g., Table 36 of the MI300 ISA manual).
Work from another wavefront (WF) in the same work group (WG) or WFs from a different WG do not have these requirements, but if they require an MCE for their instructions and all MCEs on the CU are currently busy, they must wait, and the GPU will not schedule that WF until the hardware resources are available (i.e., the scoreboard will prevent MFMAs using the same SIMD unit from being scheduled concurrently).
Although we suspect real hardware MCE implementations have multi-stage pipelines (similar to NVIDIA~\cite{Choquette2023-hopper} and RISC-V~\cite{NadaSarda2025-vortexTensorCores} GPUs), the AMD compiler appears to behave as if MFMA instructions from a given WF cannot be pipelined in MCEs when adding independent instructions or NOPs after MFMA instructions in a given thread.
Thus, we model this behavior in gem5 and focus on parallelism across WFs and MCEs within a CU.
In the future, if this support changes, the gem5 MCE code can be easily changed to support pipelining MCEs.

\subsection{Differences Between MI200 \& MI300 Support}
\label{subsec:des-200300}

Although both MI200- and MI300-class GPUs support MCEs and have MFMA instructions, AMD modified their support for matrix operations between these two GPU generations.
For example, AMD added new MFMA instructions in MI300, including a 2-block variant of \texttt{v\_mfma\_f32\_32x32x4\_bf16} which takes the same number of cycles as the 1-block variant from MI200.\footnote{In MCEs block is similar to batch size -- it represents how many blocks within the MFMA instruction the MCE can concurrently process.  Thus, an $i$-block variant means the MCE can process $i$ MFMA blocks from the instruction concurrently.}
AMD also removed others; %
of the instructions benchmarked, \texttt{v\_mfma\_i32\_16x16x16i8} was one such instruction removed in the new architecture, which we used in our MI200 benchmarks (Section~\ref{sec:meth}).
Moreover, MI300 GPUs have improved the latency (i.e., reduced the number of cycles) required to execute some MFMA instructions compared to MI200 GPUs.
This can be seen by comparing the \textbf{Expected} column in Tables~\ref{tab:hardware_nmfma-mi200} (MI200) and \ref{tab:hardware_nmfma-mi300} (MI300) for the MFMA instructions that are supported in both GPUs (e.g., \texttt{fp32\_16x16x16fp16}).
In the gem5 codebase, a comparison of supported and removed instructions, along with their cycle count, can be seen in the \texttt{mfma\_cycles} lookup table located in \texttt{src/gpu-compute/compute\_unit.cc}.

\section{Methodology}
\label{sec:meth}

\subsection{System Setup}
\label{subsec:meth-sys}

\begin{table}[tb!]
  \centering
  \vspace{1ex}
  {\footnotesize
  \begin{tabular}{|c|c|}
    \hline
    \textbf{GPU Feature} & \textbf{Configuration}\\
    \hline
    GPU Clock & 1801 MHz\\ \hline
    Total CUs & 60\\ \hline
    Num SIMD units/CU & 4\\ \hline
    Max WF/SIMD unit & 10\\ \hline
    Vector/Scalar Reg. File Size / CU & 256/12.5 KB\\ \hline
    LI Instruction Cache / 4 CU & 16 KB, 64B line, 8-way\\ \hline
    LI Scalar Cache / 4 CU & 16 KB, 64B line, 8-way\\ \hline
    L1 Data Cache / CU & 16 KB, 64B line, 16-way\\ \hline
    L1 Instruction Cache Latency & 40 cycles\\ \hline
    L1 Data Cache Latency & 140 cycles\\ \hline
    L1 Scalar Latency & 41 cycles\\ \hline
    LDS Size / CU & 64 KB\\ \hline
    LDS Latency & 65 cycles\\ \hline
    L2 Cache & 8 MB, 64B line, 32-way (16 banks)\\ \hline
    L2 Latency & 269 cycles\\ \hline
    L2 Write Policy & Write-back with write allocate\\ \hline
    Main Memory & 16 GB HBM2, 4H stacks\\ & 1000 MHz, 64 banks\\ \hline
    Main Memory Latency & 483 cycles\\ \hline
  \end{tabular}
}

  \caption{Simulated baseline GPU parameters.}
  \label{tab:config}
\end{table}

\subsubsection{gem5}
\label{subsubsec:meth-sys-gem5}

Similar to prior work, we evaluate the MFMA instructions using a tightly coupled CPU-GPU architecture with a unified address space with shared memory and coherence caches~\cite{DalmiaShashiKumar2024-cpelide}.
All CPU cores and GPU CUs are connected via a shared, inclusive L3, which also serves as the directory.
We use ROCm 6.2.2~\cite{rocm} and gem5 v24.1~\cite{binkert2011gem5, LowePowerAhmad2020-gem520}, 
We configured this gem5 setup to mimic the real MI200 and MI300 GPUs (Section~\ref{subsubsec:meth-sys-realGPUs}) as closely as possible, using GPUFS mode.
Table~\ref{tab:config} summarizes the common key system parameters, which our group previously validated to tune gem5 relative to real hardware~\cite{JamiesonChandrashekar2022-gap, RamadasKouchekinia2023-gap, RamadasKouchekinia2024-gap}.

\subsubsection{Real GPUs}
\label{subsubsec:meth-sys-realGPUs}

We compared the timing behavior of our new MFMA support in gem5 on the applications in Section~\ref{subsec:meth-bmks} using AMD Instinct MI210\textsuperscript{\texttrademark}~\cite{LohSchulte2023-mi250, amd-cdna2} MI300\textsuperscript{\texttrademark}~\cite{SmithLoh2024-mi300A, amd-cdna3} GPUs on the AMD AI \& HPC Fund~\cite{amd-hpcfund}.
For each benchmark we ensured consistent results by running each test at least five times and averaging the results.

\subsection{Benchmarks}
\label{subsec:meth-bmks}

Although our group recently added support for running large-scale ML workloads in PyTorch and TensorFlow in gem5~\cite{RamadasPoremba2024-gem5GPUFS, RamadasSinclair2024-gem5GPUFS}, in this work we focus on evaluating our MFMA support using targeted microbenchmarks because they help us test and validate their behavior in isolation, unlike large-scale workloads where MFMA operations are intermixed with numerous other instructions.
Nevertheless, our MFMA support also works for these larger workloads. 
To evaluate our added support for MI200 and MI300 GPUs, we designed a series of microbenchmarks based on tests from AMD's lab notes~\cite{amdLabNotes}.
Tables~\ref{tab:hardware_nmfma-mi200} and \ref{tab:gem5_nmfma-mi200}, summarize the MI200-class GPU microbenchmarks, while Tables~\ref{tab:hardware_nmfma-mi300}, and \ref{tab:gem5_nmfma-mi300} summarize the MI300-class GPU microbenchmarks.
We selected these MFMA instructions because we found popular PyTorch and TensorFlow ML workloads from GPT~\cite{BrownMann2020-gpt3, RadfordWu2019-gpt2, openai2024gpt4technicalreport} and MLPerf~\cite{Reddi2020mlperf-Infer, MattsonCheng2020-mlperfTrain} use them and because they demonstrate our gem5 support works for MFMA instructions of various types and precisions.

\subsection{Validation}
\label{subsec:meth-valid}

To validate that our gem5 MCE implementation provides high fidelity for MFMA instructions relative to real MI200 and MI300 GPUs, we initially attempted to time the MFMA instructions in the microbenchmarks from Section~\ref{subsec:meth-bmks}.
However, we found that AMD's HIP compiler would perform heavy software pipelining of these instructions, making timing and validating the MFMA instructions in isolation challenging.
Thus, we designed a series of handwritten, inlined assembly microbenchmarks that interleave a given MFMA instructions with GPU timing instructions (\texttt{s\_memtime}, which returns a per CU 64-bit counter value representing the clock cycle that the instruction occurred).
As discussed in Section~\ref{sec:des}, the GPU's internal scoreboard prevents MFMAs on the same SIMD unit from being executed concurrently.
Thus, by interleaving MFMA instructions and timing instructions, we can accurately time how long each individual MFMA takes in both real hardware and gem5.

\begin{lstlisting}[language=C++, label=code:mfma-validate, caption={Example AMD GPU code snippet used to validate MFMA behavior in gem5.}, float=tb, escapeinside={(*}{*)}]
  asm volatile(
      "s_waitcnt lgkmcnt(0) & vmcnt(0)\n\t" (*\label{code:pad_loc}*)
      "s_memtime %
      "s_waitcnt lgkmcnt(0)\n\t"
      "v_mfma_f32_4x4x1f32 %
      "v_mfma_f32_4x4x1f32 %
      "v_mfma_f32_4x4x1f32 %
      "v_mfma_f32_4x4x1f32 %
      "s_memtime %
      "s_waitcnt lgkmcnt(0)\n\t"
    : [start] "=r"(start), [end] "=r"(end), [D] "=v"(d)
    : [A] "v"(a), [B] "v"(b), [C] "v"(d));
  total += end - start; (*\label{code:total}*)
\end{lstlisting}

Listing~\ref{code:mfma-validate} shows an example of our microbenchmarks.
In modern AMD GPUs the \texttt{s\_memtime} instruction accesses the current clock cycle in the scalar cache.
Since the scalar pipeline is independent of the MCE pipeline (Section~\ref{sec:des}), the \texttt{s\_memtime} instruction is not guaranteed to wait for a preceding MFMA instruction to complete -- e.g., if there is only a single MFMA instruction between \texttt{s\_memtime} instructions.
To overcome this, in each test we insert multiple MFMA instructions back-to-back, with data dependencies between them.
This ensures that the MFMA instructions must complete sequentially, which in turn forces the second \texttt{s\_memtime} instruction to wait (the GPU WF scheduler will stop scheduling subsequent instructions in a WF if there are true data dependencies~\cite{GutierrezBeckmann2018-amdGem5}).
In turn this makes it easier to time the MFMA instructions on real GPUs.
Note that since we intentionally ignore the GPUs requirements for putting sufficient independent work between MFMA instructions (Section~\ref{sec:des}) in these microbenchmarks, their functional output may be incorrect.
AMD's lab notes~\cite{amdLabNotes}, which these microbenchmarks are based on, retains the compiler-inserted NOPs between MFMA instructions and validates that each MFMA instruction outputs the correct functional answers.

In Listing~\ref{code:mfma-validate}, 4 MFMAs are inlined one after another, but only 2 are needed -- since the \texttt{s\_memtime} does not wait for the final MFMA to complete given there is not a data dependence.
To calculate the time each MFMA takes, we use Equation~\ref{eq:mfma_timing}:

\begin{equation}
  \label{eq:mfma_timing}
  T_{MFMA} = (T_{total}-T_{memtime}-T_{inst})/(N_{MFMA}-1)
\end{equation}

where $T_{total}$ is \verb|total| from line~\ref{code:total}, $T_{memtime} = 40$ and $T_{inst} = 4$ were determined from microbenchmarks in prior work~\cite{JamiesonChandrashekar2022-gap, RamadasKouchekinia2023-gap, RamadasKouchekinia2024-gap}, and $N_{MFMA} = 4$ in our example kernel above. 
$T_{memtime}+T_{inst}$ is the time for the final MFMA to execute, so they are subtracted from $T_{total}$.
Since this includes the final MFMA, we also subtract 1 from $N_{MFMA}$ when dividing the total time to find individual cycle time (which the \texttt{s\_memtime} does not accurately measure, as discussed above).

\section{Results: MFMA Implementation Validation}
\label{sec:valid}

\subsection{Modeling Matrix Core}
\label{subsec:valid-model}

For each MFMA instruction, we test two to five MFMA instructions back-to-back (Section~\ref{subsec:meth-valid}) to ensure our support provides the expected behavior for various use cases on both MI200 and MI300 GPUs.
Then, we compared the timed outputs from gem5 for each GPU model to both real hardware and the corresponding tables in the ISA manuals (e.g., Table 27 in the MI300 ISA manual~\cite{mi300_isa}).
Overall, these results show that we provide accurate timing models for various MI200 and MI300 MFMA instructions, and we have incorporated these timing models for each tested MFMA instruction into gem5's mainline public support.

\begin{table}[tb!]
  \centering
  \vspace{1ex}
  {\footnotesize
  \begin{tabular}{|c|c|c|c|c|c|}
    \hline
    \multirow{2}{*}{\textbf{MFMA}} & \multicolumn{4}{c|}{\textbf{$N_{MFMA}$}} & \multirow{2}{*}{\textbf{Expected}} \\ \cline{2-5}
     & \textbf{2} & \textbf{3} & \textbf{4} & \textbf{5} & \textbf{} \\ \hline \hline
    fp64\_16x16x4fp64 & 32 & 32 & 32 & 32 & 32 \\ \hline
    fp32\_4x4x1fp32 & 8 & 8 & 8 & 8 & 8 \\ \hline
    fp32\_16x16x4fp32 & 32 & 32 & 32 & 32 & 32 \\ \hline
    \textcolor{blue}{fp32\_16x16x16fp16} & 32 & 32 & 32 & 32 & 32 \\ \hline
    \textcolor{blue}{i32\_16x16x16i8} & 32 & 32 & 32 & 32 & 32 \\ \hline
    fp64\_4x4x4fp64 & 16 & 16 & 16 & 16 & 16 \\ \hline
    \textcolor{blue}{fp32\_4x4x4fp16} & 8 & 8 & 8 & 8 & 8 \\ \hline
  \end{tabular}
}

  \vspace{1ex}
  \caption{Real Hardware MI200 latency.  Instructions in blue required padding to be accurately measured.}
  \label{tab:hardware_nmfma-mi200}
\end{table}

\begin{table}[tb!]
  \centering
  \vspace{1ex}
  {\footnotesize
  \begin{tabular}{|c|c|c|c|c|c|}
    \hline
    \multirow{2}{*}{\textbf{MFMA}} & \multicolumn{4}{c|}{\textbf{$N_{MFMA}$}} & \multirow{2}{*}{\textbf{Expected}} \\ \cline{2-5}
     & \textbf{2} & \textbf{3} & \textbf{4} & \textbf{5} & \textbf{} \\ \hline \hline
    fp64\_16x16x4fp64 & 32 & 31.75 & 31.83 & 32.06 & 32 \\ \hline  %
    fp32\_4x4x1fp32 & 8.25 & 7.88 & 7.92 & 7.94 & 8 \\ \hline  %
    fp32\_16x16x4fp32 & 32.25 & 32.5 & 32 & 32.125 & 32 \\ \hline  %
    \textcolor{blue}{fp32\_16x16x16fp16} & 32 & 31.5 & 32.33 & 32 & 32 \\ \hline  %
    \textcolor{blue}{i32\_16x16x16i8} & 32 & 32 & 32.33 & 32 & 32 \\ \hline  %
    fp64\_4x4x4fp64 & 16 & 16 & 16.33 & 16.25 & 16 \\ \hline  %
    \textcolor{blue}{fp32\_4x4x4fp16} & 7 & 7.5 & 7.67 & 8 & 8 \\ \hline  %
  \end{tabular}
}

  \vspace{1ex}
  \caption{MI200 latency in gem5.  Instructions in blue required padding to be accurately measured.}
  \label{tab:gem5_nmfma-mi200}
\end{table}

\begin{table}[tb!]
  \centering
  \vspace{1ex}
  {\footnotesize
  \begin{tabular}{|c|c|c|c|c|c|}
    \hline
    \multirow{2}{*}{\textbf{MFMA}} & \multicolumn{4}{c|}{\textbf{$N_{MFMA}$}} & \multirow{2}{*}{\textbf{Expected}} \\ \cline{2-5}
     & \textbf{2} & \textbf{3} & \textbf{4} & \textbf{5} & \textbf{} \\ \hline \hline
    fp64\_16x16x4fp64 & 32 & 32 & 32 & 32 & 32 \\ \hline
    fp32\_4x4x1fp32 & 8 & 8 & 8 & 8 & 8 \\ \hline
    fp32\_16x16x4fp32 & 32 & 32 & 32 & 32 & 32 \\ \hline
    \textcolor{blue}{fp32\_16x16x16fp16} & 16 & 16 & 16 & 16 & 16 \\ \hline
    fp64\_4x4x4fp64 & 16 & 16 & 16 & 16 & 16 \\ \hline
    \textcolor{blue}{fp32\_4x4x4fp16} & 8 & 8 & 8 & 8 & 8 \\ \hline
    
  \end{tabular}
}

  \caption{Real Hardware MI300 latency. Instructions in blue required padding to be accurately measured.}
  \label{tab:hardware_nmfma-mi300}
\end{table}

\begin{table}[tb!]
  \centering
  \vspace{1ex}
  {\footnotesize
  \begin{tabular}{|c|c|c|c|c|c|}
    \hline
    \multirow{2}{*}{\textbf{MFMA}} & \multicolumn{4}{c|}{\textbf{$N_{MFMA}$}} & \multirow{2}{*}{\textbf{Expected}} \\ \cline{2-5}
     & \textbf{2} & \textbf{3} & \textbf{4} & \textbf{5} & \textbf{} \\ \hline \hline
    fp64\_16x16x4fp64 & 32 & 32.13 & 32.08 & 31.94 & 32 \\ \hline %
    fp32\_4x4x1fp32 & 7.5 & 8 & 7.92 & 8.06 & 8 \\ \hline %
    fp32\_16x16x4fp32 & 32 & 31.75 & 31.83 & 32 & 32 \\ \hline %
    \textcolor{blue}{fp32\_16x16x16fp16} & 15 & 16.5 & 16.33 & 16.25 & 16 \\ \hline %
    fp64\_4x4x4fp64 & 16 & 16 & 16 & 16.25 & 16 \\ \hline %
    \textcolor{blue}{fp32\_4x4x4fp16} & 8 & 8 & 7.67 & 8.25 & 8 \\ \hline %
  \end{tabular}
}

  \vspace{1ex}
  \caption{MI300 latency in gem5. Instructions in blue required padding to be accurately measured.}
  \label{tab:gem5_nmfma-mi300}
\end{table}

\subsubsection{MI200}
\label{subsubsec:valid-model-mi200}

Tables~\ref{tab:hardware_nmfma-mi200} and \ref{tab:gem5_nmfma-mi200} compare the MFMA latency on real MI200 GPUs and gem5's simulated MI200 model respectively.
Overall, our gem5 MFMA implementations show very high fidelity relative to real MI200 GPUs: 1.5\% mean absolute percentage error (MAPE) 
relative to real MI200 GPUs across 2-5 MFMAs.
Although all number of back-to-back MFMAs have low MAPE values, in general we find that as the number of MFMAs increase, the MAPE relative to the real MI200 GPUs decreases from 2.3\% MAPE for 2 MFMAs to 0.4\% for 5 MFMAs.
We believe this happens because additional MFMAs lead to longer sections between timing measurements -- reducing the likelihood that transient effects impact the timing results.
However, for some MFMA instructions (shown in blue) additional care is required to obtain these results.
Specifically, for these MFMA instructions we found that padding was required to get accurate measurements.
To do this, we added multiple \texttt{s\_nop} instructions before line~\ref{code:pad_loc} of Listing~\ref{code:mfma-validate}.
Padding prevents the GPU from needing to fetch a new instruction cache cacheline in the middle of our sequence of instructions that are timing the MFMA latency.
We discuss the implications of this padding further in Section~\ref{sec:disc}.

\subsubsection{MI300}
\label{subsubsec:valid-model-mi300}

Tables~\ref{tab:hardware_nmfma-mi300} and \ref{tab:gem5_nmfma-mi300} similarly compare the MFMA latency on real MI300 GPUs and gem5's simulated MI300 model.
Similar to the MI200 results, our gem5 MFMA implementations again provide high fidelity relative to real MI300 GPUs: 1.3\% MAPE.
Once again, microbenchmarks with more MFMA instructions back-to-back tend to exhibit lower MAPE than those with fewer MFMA instructions: 1.2\% MAPE for 5 MFMAs versus 2.1\% MAPE for 2 MFMAs.
However, in some situations we find that obtaining accurate latency measurements is difficult and can vary, even after padding the appropriate MFMA instructions to avoid unintended instruction cache misses during the timing section.
Here, we find that the variability occurs when gem5 runs the CPU portion in KVM mode, which is non-deterministic.
Nevertheless, these differences are very small, and in general these results show that our gem5 MFMA implementations are well correlated with real MI300 GPUs.

\subsection{What-if Analysis}
\label{subsec:valid-whatif}

To further demonstrate the benefits of our MCE and MFMA additions to gem5, we added a new configuration parameter: \verb|--mfma-scale|.
By multiplying the default latency for a given MFMA instruction (Section~\ref{sec:des}) by this configuration parameter, users can increase or decrease the number of cycles a MFMA takes in a MCE.
Thus, this parameter allows users to conduct what-if analysis to examine how potential improvements to the MCEs affect an application's performance.
Specifically, here we use our microbenchmarks with 2 MFMA instructions back-to-back (Listing~\ref{code:mfma-validate}).

Table~\ref{tab:gem5_scale-mi300} shows the effect of doubling MFMA latency (by setting \texttt{--mfma-scale=2}) in gem5's simulated MI300 model.
Intuitively, since our microbenchmarks put 2 MFMA instructions back-to-back, doubling MFMA latency should cause each microbenchmark to take twice as long compared to the default of \texttt{--mfma-scale=1}.
In general, the results in Table~\ref{tab:gem5_scale-mi300} show that this is the case.
However, not all measurements show perfect, linear scaling: overall we obtain a 3\% MAPE relative to doubling the latencies.
This happens for similar reasons to Section~\ref{subsec:valid-model}: non-determinism from KVM.
Nevertheless, these results further demonstrate the validity of our gem5 MFMA implementation and show how changes to \texttt{--mfma-scale} affect $T_{MFMA}$.

However, it is important to note that we have complete control over the instructions being executed and measured in our microbenchmarks, due to our use of inlined assembly.
This allowed us to observe near perfect, linear scaling in our MFMA what-if analysis.
In real workloads the compiler is responsible for enforcing timing conditions, making it more difficult to remove NOP instructions as MFMA latency scales.
Thus, it is important to also evaluate the impact of this parameter on a real workload.
We discuss the implications of this further in Section~\ref{sec:disc}.

\begin{table}[tb!]
  \centering
  \vspace{1ex}
  {\footnotesize
  \begin{tabular}{|c|c|c|c|c|c|}
    \hline
    \multirow{2}{*}{\textbf{MFMA}} & \multicolumn{2}{c|}{\textbf{\texttt{--mfma-scale}}} \\ \cline{2-3}
     & \textbf{1} & \textbf{2}\\ \hline \hline
    fp64\_16x16x4fp64 & 32 & 63 \\ \hline
    fp32\_4x4x1fp32 & 7.5 & 16 \\ \hline
    fp32\_16x16x4fp32 & 32 & 65 \\ \hline
    \textcolor{blue}{fp32\_16x16x16fp16} & 15 & 33 \\ \hline
    fp64\_4x4x4fp64 & 16 & 32 \\ \hline
    \textcolor{blue}{fp32\_4x4x4fp16} & 8 & 16 \\ \hline
    
  \end{tabular}
}

  \vspace{1ex}
  \caption{MI300 scaled latency in gem5 with $N_{MFMA}=2$. Instructions in blue required padding to be accurately measured.}
  \label{tab:gem5_scale-mi300}
\end{table}

\section{Implementation Limitations}
\label{sec:disc}

\noindent
\textbf{Removing NOPs}: As discussed in Sections~\ref{sec:des} and \ref{sec:valid}, AMD's compiler-based approach to enforcing correctness with MFMA instructions introduces challenges when performing what-if analysis in simulators such as gem5.
Consequently, scaling the latency of MFMA instructions in gem5 without corresponding changes to the compiler to change how much independent work must be inserted between MFMA instructions in the same WF will likely not result in linear scaling like we observed in Section~\ref{subsec:valid-whatif}.
This is a limitation of our MCE support in gem5.
To overcome this, users have two options: they can either a) write inlined GPU assembly snippets that reduce the amount of independent work between MFMAs to mirror the changes in \verb|--mfma-scale|, and then use these snippets in larger workloads or b) rewrite the compiler to adjust how much independent work must be placed between MFMA instructions in the same WF based on the desired MFMA scaling factor.
Crucially, AMD's ROCm~\cite{rocm} GPU stack is open source, enabling sufficiently motivated researchers to make the necessary changes to the compiler.
Obviously, both of these options also have downsides (e.g., requiring compiler changes or inlined assembly), but regardless it is important that users understand the implications of how AMD enforces correctness for MFMA instructions.

\noindent
\textbf{MFMA Instruction Failures}: AMD recently introduced a new access mode: \texttt{s\_set\_gpr\_idx}, which utilizes a separate (GPR) addressing mode to improve the performance of specific code sequences.
However, currently this addressing mode is unsupported in gem5.
As a result, some MFMA instructions (e.g., \texttt{v\_mfma\_fp32\_32x32x8fp16} and \texttt{v\_mfma\_fp32\_32x32x1fp32}) are unsupported.
Although it is possible to implement this addressing mode, it requires updating every AMD GPU instruction in gem5 to check if it is using this addressing mode.
Since there are thousands of AMD GPU instructions, and relatively few MFMA instructions use this addressing mode, we instead focused on adding support for the MFMA instructions that do not use this addressing mode.

\noindent
\textbf{Cache-Line Aligned MFMA Instruction Sequences}: As discussed in Section~\ref{sec:valid}, the timing of some MFMA tests is sensitive to fetching new cache lines in the middle of our timed sequence of instructions.
Without cache line aligning these tests, the gem5 and real GPU behavior often exhibit unexpected behavior.
However, since we observed that both gem5 and the real GPU exhibit similar unexpected behavior when an additional cache line is fetched in the middle of the region of interest, we do not believe changes are required to gem5.
Instead, if users observe unexpected results with specific numbers of MFMA instructions and this significantly impacts the overall behavior of the workload, it is recommended they investigate the cache line alignment of the region of interest.

\section{Conclusion}
\label{sec:conc}

As ML workloads increasingly drive the requirements of future systems, it is imperative that state-of-the-art tools like gem5 evolve with them.
Without significant improvements, researchers will be unable to perform early stage co-design for these important workloads while being confident the results are representative.
In particular, a major shortcoming of gem5's support for ML workloads is its lack of support for MCEs.
In this work, we rectify this issue and add support for MCEs and their MFMA instructions in modern MI200 and MI300 AMD GPUs in gem5.
Our results show that our support provides high fidelity MFMA support for both MI200 and MI300 GPUs.
For a wide variety of MFMA instructions, our gem5 support provides identical or near identical latencies: 1.5\% and 1.3\% MAPE relative to real MI200 and MI300 GPUs, respectively.
Moreover, our results also show how researchers can leverage this support to perform what-if analysis by scaling the performance of MFMA operations in gem5.
Overall, this work is a significant step forward and enables researchers to perform rapid, high fidelity experimentation for a variety of modern, important GPU workloads.

\section*{Acknowledgments}

This work is supported in by the National Science Foundation CSSI grant Frameworks-2311889.
We also thank Matt Poremba at AMD Research for his advice, which greatly improved the quality of this work.

\ifCLASSOPTIONcaptionsoff
  \newpage
\fi

\bibliographystyle{IEEEtran}
\bibliography{references}

\end{document}